\documentstyle[11pt]{article}

\newcommand{\beq}{\begin{equation}}
\newcommand{\eeq}{\end{equation}}
\newcommand{\beqa}{\begin{eqnarray}}
\newcommand{\eeqa}{\end{eqnarray}}

\begin{document}
\begin{center}
{\bf   Results in supersymmetric field theory\\
\vspace{0.5cm}
from 3-brane probe in F-theory}\\
\vspace{1cm}
Ansar Fayyazuddin
\\
The Niels Bohr Institute, Copenhagen University, Blegdamsvej 17, DK-2100 Copenhagen {\O}, Denmark
\end{center}

\vspace{.25cm}
\begin{abstract}
Employing Sen's picture of BPS states on a 3-brane probe world volume
field theory in a F-theory background, we determine some selection rules 
for the allowed spectrum in 
massless $N_{f}\leq 4$ $SU(2)$ Seiberg-Witten theory.  The
spectrum for any $N_f \leq 4$ is consistent with previous conjectures
and analysis.
\end{abstract}
\vspace{0.25cm}

NBI-HE-97-03

January 1997

\vspace{0.50cm}
Recent progress in string theory has as a by product yielded new results
in supersymmetric field theory.  While some of these results have been 
known and are to some extent a test of string theory ideas, others
are new field theory results.  One notable success \cite{witten}
is the explanation
of the SL(2,Z) symmetry of $N=4$ Yang-Mills.  The field theory SL(2,Z)
is realized as a geometric action on a compactifying manifold of the
underlying string theory.  Geometrizing field theory 
 symmetries by embedding them in string theory 
make non-perturbative results in field theory, which are basically
inaccessible by means of standard techniques, transparent geometric facts 
from the string perspective. 
The purpose of this paper is to demonstrate how the spectrum of 
$N=2$ supersymmetric QCD with gauge group $SU(2)$ can be found by probing
a certain F-theory background proposed by Sen.  

F-theory\cite{vafa} is a powerful method of constructing type IIB vacua.  
By definition,
F-theory on a manifold which admits an elliptic fibration is a type IIB 
compactification on the base manifold of the fibration with the field  
$\tau = A_0 + ie^{-\phi}$, ($A_0$ is the Ramond-Ramond zero form and 
$\phi$ the dilaton) idenified with the modular parameter (restricted
to lie in the fundamental domain) of the elliptic fiber.   Thus the
type IIB compactification has in general some background 7-branes and
D-instantons turned
on.  Sen \cite{senorient} studied the compactification of F-theory on K3 in the
$T^4 /Z_2$ orbifold limit.  In this case the $\tau$ parameter of the
elliptic fiber does
not vary over the base manifold ($T^2 /Z_2$) and is a conventional IIB 
compactification on a $T^2$
orientifold with four 7-Dbranes turned on at the location of each of the 
four orientifold fixed planes to cancel the orientifold charge.
Perturbing  away from this point in $K3$ moduli space Sen found
that the IIB background ceased to make sense and had to be modified.
Viewing the IIB compactification as coming from F-thoery allowed Sen
to specify the modification.  F-theory required that there should
really be six 7-branes located close to each of the orbifold points
as you move away from the orbifold limit of K3.  Thus F-theory predicts
that each IIB orientifold fixed plane should split into two 7-branes which
are not D-branes with respect to the same $(p,q)$ string.  Exploiting
the fact that a similar splitting of a classical singularity occurs
in the $SU(2)$ Seiberg-Witten theory with four hypermultiplets
when the masses of the hypermultiplets are allowed to become different
from zero, Sen was able to parameterize the K3 (of F-theory
compactification) locally close
to the orbifold limit by a curve identical to the one for the Seiberg-Witten
theory.   

The reason why Seiberg-Witten  mathematics arises in this context
was explained by Banks et. al\cite{banks}.  They argued that there is 
essentially
a unique probe that one can introduce in this 7-brane background which
does not modify the background geometry.  This probe is a 3-brane.  It
has a $U(1)$ gauge theory living on its world-volume which becomes
enhanced to $SU(2)$ when it approaches the orientifold fixed plane.
The hypermultiplet fields living on the world volume theory are strings 
stretched between
the 3-brane and the various 7-branes.  One can account for all
the global symmetries of the Seiberg-Witten theory as  gauge symmetries
from the point of view of the 7-brane background.  The vector multiplets
correspond to strings starting and ending on the 3-brane probe.
Sen showed that for the states on the 3-brane probe to saturate a
BPS bound there must be a geodesic path connecting the 3-brane to
the 7-branes (for the hypermultiplets) and to the 3-brane probe itself
(for vector multiplets).  This criterion is a minimal test for the
existence of states, since it only requires that the BPS bound be
saturated.

Using D-brane probes to find results in field theory is a very powerful
method.  It has recently been used in \cite{banks}-\cite{dls}
to gain a wealth of information about field theories in $6,5,4,$ and $3$
dimensions.  

We now briefly review the situation we will be studying to set the
notation and indicate the strategy.  Consider a K3 which has an elliptic
fibration:

\beq
y^2 = x^3 + f(z)x + g(z)
\eeq 
This equation describes a K3 if $f,g$ are of degree
eight and twelve respectively in $z$.  This can easily be shown by
introducing a new coordinate $v$ to make the equation homogeneous in 
weighted projective space $WP_{1,1,4,6}$
\beq
W = -y^2 + x^3 + \alpha\prod_{i=1}^{8} (z-vw_{i})x + 
\prod_{i=1}^{12}(z-vw'_{i}) 
\eeq
where $x,y,z,v$ have weights $4,6,1,1$ respectively, and 
$f\equiv \alpha\prod_{i=1}^{8} (z-vw_{i})$, 
$g\equiv\prod_{i=1}^{12} (z-vw'_{i})$ (using the freedom to rescale $y,x$ to
set the coefficient of the highest order term in $g$ to 1).
By setting $v=1$ one recovers the original equation.  
Sen \cite{senorient} studied the $T^{4}/Z_2$ orbifold limit of the 
$K3$ in which $f^3/g^2$
is a constant.  In this limit the modular parameter $\tau$ of the
elliptic fiber does not vary as a function of $z$ and Sen showed
that it described a type IIB orientifold in which the Ramond-Ramond
and (Neveu-Schwarz)$^2$ two forms are modded out.  In the type IIB description
the orientifold carries $-4$ units of 7-brane charge which has to be
locally cancelled by $4$ Dirichlet 7-branes so that $\tau$
does not vary.  When one tries to move 
the four 7-branes away from the orientifold fixed plane one finds that
$\tau$ develops a negative imaginary part and thus quantum corrections
have to modify the background.  Viewing it as a F-theory background
instead one sees that the there are instead six 7-branes at the fixed
points of the $Z_2$ and moving away from this point in K3 moduli space
splits the IIB orientifold plane into two 7-branes which are not Dirichlet
branes of the same $(p,q)$ string.   This gives a non-perturbative
description of the IIB background.

Focusing ones attention to the neighborhood of one of the 
orientifold fixed planes
(i.e. working on $R^2/Z_2$) one can parameterize the $K3$ locally by
the Seiberg-Witten curve for $SU(2)$ Yang-Mills coupled to 
$N_f =4$ hypermultiplets.  The deformations away from the orbifold
limit of $K3$ can be parameterized in terms of the four hypermultiplet masses
which when set to zero give the orbifold limit.
The curve is given by:
\beq
y^2 = x^3 + {\tilde f}x + {\tilde g}
\eeq
with $\tilde f$ of degree 2 and $\tilde g$ of degree 3 in $z$.    
Identifying $u$ in Seiberg-Witten theory with $z$, the coordinate of
the base manifold,  ${\tilde f}$ and $\tilde g$ are obtained by putting 
the $N_{f}=4$,
$SU(2)$ curve given in \cite{seibwit2} into Weierstrass form, and rescaling
$x,y$ to set the coefficient of the cubic term in $g$ equal to one.
The positions of the 7-branes $z_{i}$ are given through the formula:
\beq
\Delta = (27+4\alpha^{3})\prod_{i=1}^{6}(z-z_{i})
\eeq
where $\Delta$ is the discriminant of the elliptic fiber, and $\alpha$ is
the coefficient of the quadratic term in ${\tilde f}$.

One can now introduce a parallel 3-brane to probe this 7-brane background and
identify $z$ as the coordinate of the 3-brane\footnote{
Of course, to specify the position of the 3-brane one needs to specify
$6$ real coordinates, however, since the 3-brane is parallel
to the 7-branes the world volume theory does not depend on the remaining $4$
coordinates due to translational invariance in those directions.}.
A string ending on the 3-brane probe corresponds to a state in the
field theory living on the 3-branes world-volume.  As noted in{\cite{banks,
senprobe}}
a $(p,q)$ string stretched between the probe and the
background 7-branes will correspond to a hypermultiplet state of charge
$(p,q)$ with respect to the photon field corresponding to a $(1,0)$ 
fundamental string of zero length with both ends on the probe.  Such
a state will also carry global Chan-Paton factors from the point of
view of the 3-brane which are gauge Chan-Paton factors
of the 7-brane background.
The mass of $(p,q)$ hypermultiplet is given by integrating the tension
of the $(p,q)$ string with the appropriate metric over the spatial extent of 
the string.  The tension \cite{schwarz} of a $(p,q)$ string is given
by:
\beq
T_{\left( p,q\right)} = \frac{1}{\sqrt{\tau_2}}\mid p + q\tau\mid,
\eeq 
and the metric in the 7-brane background is given by\cite{greene, senorient}:
\beq
ds^2 = \tau_{2}\mid\frac{\eta\left(\tau\right)}{\sqrt{2}\eta\left(\tau_{0}\right)}
\prod_{i=1}^{6}\left( z-z_{i}\right)^{-\frac{1}{12}}dz\mid^{2}.
\eeq
$\tau_{0}$ is the asymptotic value of $\tau$.
We have also fixed the arbitrary constant factor in the metric to make
it consistent with the definition of $(a, a_{D})$ in the Seiberg-Witten
theory.
Thus the mass of a state in the 3-brane world-volume field theory
corresponding to a $(p,q)$ string stretched between the 3-brane and
a $7-brane$ is given by \cite{senprobe}
\beq
m_{\left(p,q\right)} = \int T_{\left(p,q\right)}ds
\eeq
where the integral is done over a path connecting the 3-brane probe
to the 7-brane such that it minimizes the mass. 
One can have a variety of states in the world-volume theory
corresponding to strings stretched between the same 3- and 7-branes
but differing topologically in the way in which the string winds around 
the positions 
of the 7-branes.  Since a $(p,q)$ string can only end on $(p,q)$ 7-branes
\cite{douglasli}
this is equivalent to saying that the 7-brane charge depends, from the
point of view of the probe, on the way
in which the probe approaches the 7-brane \cite{senprobe, johansen}.

The reason one minimizes the mass is to saturate the BPS bound.
If there is no geodesic minimizing the mass of a particular state
then clearly such a state cannot be a BPS state, thus the existence
of the geodesic is certainly a necessary condition for the existence
of a stable BPS state.  In \cite{klemmetal, johansen}
it was argued that the existence of this geodesic should be taken also as 
a sufficient condition of
stability.  
The purpose of this paper is to show that this criterion
restricts the allowed states to only those conjectured and/or 
proven in the literature.  Further, assuming that the existence
of the required geodesic is a sufficiently stringent criterion for the
existence of the BPS saturated state in the world volume theory allows
one to prove the $SL\left(2,Z\right)$ symmetry of the spectrum of
the $N_f =4$ theory.   

As noted in \cite{senprobe} the condition for minimizing the mass 
of a stretched string is that the path along which the string is stretched
satisfy:
\beq
\frac{\eta\left(\tau\right)}{\sqrt{2}\eta\left(\tau_{0}\right)}
\prod_{i=1}^{6}\left( z-z_{i}\right)^{-\frac{1}{12}}
\left(p +q\tau\right)\frac{dz}{dt} = \mbox{constant} \label.
\eeq
Sen then showed that   
\beq
\frac{da}{dt}=\frac{\eta\left(\tau\right)}{\sqrt{2}\eta\left(\tau_{0}\right)}
\prod_{i=1}^{6}\left( z-z_{i}\right)^{-\frac{1}{12}}\frac{dz}{dt}.
\eeq
The relation
\beq
da_{D} = \tau da
\eeq
then implies that the mass of the string stretched between a $(p,q)$ 7-brane 
and the probe 
is given by $m_{\left(p,q\right)} = \mid pa+qa_{D}\mid$.  These formulas
are valid as long as one does not go around a singularity.  If one continues
$a, a_D$ around a singularity $a, a_D$ undergo a $SL(2,Z)$ transformation
and do not return to their original value.  However,
one can work with multi-valued fields ${\tilde a}, {\tilde a}_{D}$
defined by continuing $a, a_D$ along the geodesic.  Then the mass
formula remains the same but the identity of the 7-brane at the other end 
($z(t=1)$) of the string changes to $(p',q')$ as:
\beq
(p,q) = (p',q')M.
\eeq
where the matrix $M$ is defined by:
\beq
\left(\begin{array}{c}{\tilde a}(z(1)=z_{i}) \\ {\tilde a}_{D}(z(1)=z_{i})
\end{array}\right) =
M\left(\begin{array}{c} a(z_{i}) \\ a_{D}(z_{i})\end{array}\right),
\eeq
$M\in SL(2,Z)$ is the monodromy matrix associated with the geodesic due
to winding around the positions of the 7-branes.
Such a string configuration corresponds to a $(p,q)$ state in the
3-brane world-volume theory.

In the rest of this paper we will consider the problem of whether or
not there exist geodesics corresponding to the various $(p,q)$ states
in theories with $N_f \leq 4$.  We will only treat the massless cases
in the following.  The problem is to find geodesics starting at
the position of the 3-brane probe and ending either at the position of 
a 7-brane (for hypermultiplets) or the 
3-brane itself (for vector multiplets). 

The simplest case from the point of view of 
F-theory is also the most difficult from the point of view of 
field theory and illustrates nicely the power of this approach.
The massless $N_{f}=4$ theory is a conformally invariant theory
with a non-running coupling constant $\tau = \tau_{0}$.
In the F-theory background all the 7-branes are located at the origin ($z=0$).
The geodesic equation is simply:
\beq
z^{-\frac{1}{2}}\left( p+q\tau_{0}\right)\frac{dz}{dt} = \mbox{constant}
\eeq   
We look for solutions with $z\left(0\right) = u$ and $z\left(1\right) = 0$.
Since the orientifold fixed plane $z=0$ is $SL(2,Z)$ invariant any
$(p,q)$ string is allowed to end on it\cite{douglasli}.  
These solutions correspond to hypermultiplets with electric charge
$p$ and magnetic charge $q$.
The solution is simply given by $z= u\left(1-t\right)^2$.
Since for all relatively prime $p,q$ there is a $(p,q)$ string,
this shows that the necessary geodesics for the $(p,q)$ hypermultiplets
exist.  In \cite{seibwit2} the existence of a set of  vector multiplets 
was also predicted, they carry charge $(2p,2q)$
for $p,q$ relatively prime.  In the world volume theory 
the vector multiplets correspond to strings starting and ending on the
3-brane: $z\left(0\right) = z\left(1\right) = u$.  
The solution is given simply by  $z= u\left(1-2t\right)^2$.
It is satisfying to note that these are the only possibilities.
There are no geodesics corresponding to states with charge $(np,nq)$
with $n>2$.   If we accept the criterion for existence of BPS states
proposed in \cite{klemmetal, johansen} that the existence of a geodesic
implies the existence of the corresponding state in the world volume theory
then we have proven the $SL(2,Z)$ symmetry conjectured
in \cite{seibwit2}.  The $SL(2,Z)$ symmetry of the $N_{f}=4$ has recently
been proven using different methods by Ganor, Morrison and 
Seiberg in \cite{ganor}.

To study the cases $N_f <4$ we have to move $4-N_f$ D7-branes
out to infinity while keeping the distances between the remaining
7-branes non-zero.  This is achieved by taking
\beq
\lim_{q->0, m_{i}->\infty}q^{\frac{1}{2}}\prod_{i=N_{f}+1}^{4}m_{i}
= \Lambda_{N_{f}}^{4-N_{f}} = \mbox{constant}
\eeq 
where $q=\exp i2\pi\tau_0$.  We also tune the remaining parameters
to put as many mutually local 7-branes at the same position as possible.
This corresponds to putting the bare masses in the world volume theory
to zero to gain a maximal global symmetry
or a maximal gauge group in the background theory.
The metrics for the $N_{f}<4$ cases with maximal flavor symmetry are: 
\beq
ds^2 = \tau_{2}\mid\frac{\eta\left(\tau\right)\Lambda^{(N_{f}-4)/6}}
{\sqrt{2}}
\prod_{i=1}^{N_{f}+2}\left( z-z_{i}\right)^{-\frac{1}{12}}dz\mid^{2}.
\eeq
For $N_{f}=3$ we have:
\beq
z_{i}=0, \mbox{   $i=1,...,4$ and } z_{5} = 16\Lambda^2
\eeq
giving a total gauge symmetry of $U(1)\times SU(4)$ in the 7-brane world volume
theory.  For $N_{f}=2$ we have
\beq
z_{1,2}= 8\Lambda^2 , z_{3,4} = -8\Lambda^2
\eeq
giving a total gauge symmetry of $U(1)\times SU(2)\times SU(2)$ while
for $N_f =1$:
\beq
z_{m}= -6\Lambda^{2}\exp\frac{i2m\pi}{3}\mbox{ $m=1,2,3$} 
\eeq
giving a total gauge symmetry of $U(1)$, finally for $N_{f}=0$:
\beq
z_{1} = 8\Lambda^{2}, z_{2} = -8\Lambda^2. 
\eeq

As is known from \cite{seibwit2, af, afs, mm} there is a curve of marginal
stability $C_{0}$ on which the ratio $a_{D}/a$ is real.  This curve
is diffeomorphic to a circle.  On this curve the
kinematic threshold for decay of BPS states into others
is achieved.  The curve seperates the regions $\mbox{Im }\frac{a_{D}}{a}<0$
and $\mbox{Im }\frac{a_{D}}{a}>0$ which we will denote by $M_{-}$ and $M_{+}$
respectively. 
The question of which states exist on either side of the
curve $C_0$ is a difficult one from the perspective of 
field theory since the physics in $M_{-}$ is strongly coupled
while in $M_{+}$ one can work in the weakly coupled region but
the calculations are rather complicated.  
Nonetheless we will see that one gets a very satisfying answer to
this question in our context.  Although it is generally difficult
to prove the existence of the appropriate geodesics short of 
doing so numerically, one can exclude the existence
of certain geodesics. Thus our goal will be modest, we will simply
try to show that only
those states which  become massless at some point in moduli space
{\em can} exist in $M_-$.  Also our analysis
will reveal that $M_+$ can
only support certain ``T-duals'' of the hypermultiplets which become
massless at some point in moduli space.

To understand how one can see this recall that the points $z_i$, which are the
locations of the 7-branes, represent values of the field
theory order parameter at which a hypermultiplet
in the 3-brane world volume theory becomes massless since a string stretched
between the 3- and 7-brane becomes of zero length at these points.  
Using the connection
above between $a_{D}, a$ and geometrical quantities in the F-theory
background, we see that all the 7-branes lie on this curve of marginal
stability.  It is crucial for what follows
that since all monodromies lie in $SL(2,Z)$ no monodromy 
can change the sign of the imaginary part of $a_{D}/a$ which transforms
exactly like $\tau$, thus it is convenient to work with the $SL(2,Z)$ 
invariant K{\" a}hler potential 
$K=\mbox{Im }a_{D}{\bar a}$ which has the same sign as $\mbox{Im }a_{D}/a$.
Suppose we have a geodesic of a $(p,q)$ string starting
(at $t=0$) at the location of the 
3-brane and ending (at $t=1$) on a $(p',q')$ 7-brane: 
\beq
p{\tilde a}(z(t)) + q{\tilde a}_{D}(z(t)) = c(1-t) 
\eeq 
for constant $c = pa(u) + qa_{D}(u)$.  It is easy to show
that for $q> 0$:
\beq
K(t)=\mbox{Im }a_{D}{\bar a} = \frac{1}{q}(1-t)
\mbox{Im }{\bar{\tilde a}}{c}.
\eeq  
The geodesic crosses the curve of marginal stability either when
$t=1$ or $\mbox{Im }{\bar c}a =0$.  Since 
\beq
\mbox{Im }{\bar c}{\dot{\tilde a}} = 
\frac{q\mid c\mid^{2}}{\mid q\tilde\tau +p\mid^2}
\mbox{Im }\tilde\tau
\eeq  
we see that $\mbox{Im }{\bar c}{\tilde a}$ is increasing along the geodesic.
This implies that geodesics emanating at a point $u\in M_{-}$
never cross the curve of marginal stability
except at its $t=1$ end point (since $K(z(0)=u)=\mbox{Im }a_{D}(u){\bar a}(u)
= \frac{1}{q}\mbox{Im }{\bar a}(z(0))c= -\frac{1}{q}\mbox{Im }{\bar c}a(z(0))
<0$).
But this implies that there are no geodesics 
which wind around the locations 
of one or more 7-branes since then they would cross the curve of marginal 
stability at least once
for $ 0<t<1$.  This implies that the only allowed geodesics correspond to
states with charge vectors equal to those of the background 7-branes 
(which from the 3-brane world volume field theory perspective are states
becoming massless somewhere in
moduli space).
In conclusion for points $u\in M_{-}$ the only allowed
hypermultiplets in the theory are those which become massless for some
value of $u$.  

For $N_{f}=0,2,3$ where there are only two singularities, the 
above analysis implies that for 
$u\in M_{+}$ the only allowed
hypermultiplets are $(p,q)= (p',q')M_{\infty}^{n}$ with $n\in Z$
where $(p',q')$ is the charge vector of a 7-brane (or, from the world
volume theory of the 3-brane perspective, the charge of a state which becomes
massless at a certain point in moduli space).  
The reason is that although a geodesic is allowed to pass through the curve
$K=0$ it may do so at most once.  But the only topologically 
non-trivial paths connecting $u\in M_{+}$ to a $z_{i}$
while crossing $C_{0}$ at most once are those which undergo $M_{\infty}$
any number of times.  The $M_{\infty}^{n}$ copies of the states becoming 
massless at some point in moduli space also inherit the global quantum
numbers of their progenitor since the string carries Chan-Paton factors
of the same group.  If for  $N_{f}=3$ the four 7-branes 
at $z=0$ are assigned charge $(0,1)$ then the states $(n,1)$ transform
as a $\mbox{\boldmath $4$}$ of $SU(4)$, while the $(1-2n, 2)$ are 
$SU(4)$ singlets.  Similarly, for $N_{f}=2$ there are states with charge
$(2n,1)$ which transform as ($\mbox{\boldmath $2$},1$) 
and charge $(1+2n,1)$ which transform as ($1$,$\mbox{\boldmath{$2$}}$) 
under the $SU(2)\times SU(2)$ global symmetry group.  It would seem
that a richer spectrum is allowed for $N_{f}=1$ in $M_{+}$ by our analysis.
We do not believe that this is the case but our methods are too elementary
to further restrict the spectrum.

The vector multiplets can similarly be analysed.  A photon vector multiplet
is a, say, $(p,q)$ string starting and ending at $z=u$ such that:
\beq
p{\tilde a}(z(t))+ q{\tilde a}_{D}(z(t)) = c
\eeq
with $c=pa(u) + qa_{D}(u)$ a constant.  Such a string has zero length
and corresponds to a neutral massless particle.
A $W$ boson is a $(p,q)$ string starting and ending at $z=u$ such that:
\beq
p{\tilde a}(z(t))+ q{\tilde a}_{D}(z(t)) = c(1-2t).
\eeq
This corresponds to a vector particle of charge $(2p,2q)$ in the
world volume theory.  The allowed vector particles is rather restricted
already at this level since 
\beq
p{\tilde a}(z(1))+ q{\tilde a}_{D}(z(1)) = 
-(p{\tilde a}(z(0))+ q{\tilde a}_{D}(z(0))).
\eeq
Thus there must exist a monodromy which acts like $-1$ on 
$p{a}(z(t))+ q{a}_{D}(z(t))$.  This condition restricts $(p,q)=(1,0)$ 
for $N_{f}<4$ with the monodromy at infinity acting as $-1$ on it.
The above analyses for hypermultiplets applies equally to the present
case showing that a geodesic emanating at a point $u\in M_{-}$ cannot
pass through $C_0$ more than once, hence disallowing geodesics 
appropriate for a 
charged vector multiplet.  If $u\in M_{+}$ then a geodesic corresponding
to a $W$-boson is allowed by our analyses.

In conclusion we have given a picture of allowed states in $N_{f}\leq 4$
massless $SU(2)$ Seiberg-Witten theory.  We have employed the condition
that states saturate the BPS bound as our sole criterion for their
existence.  This condition is powerful enough to restrict the spectrum
to only those conjectured or proven in the literature
\cite{seibwit2,af,mon,bilal,klemmetal,brand}. For $N_{f}=1$
we believe that the spectrum in $M_{+}$ is more restricted than
that allowed by our considerations.  For $N_{f}\leq 3$ we were
only able to exclude the existence of states but not to find the 
geodesics explicitly for those conjectured to exist. 

Acknowledgements
I would like to thank M. Bill\'o, P. di Vecchia, J. L. Petersen, and D. Smith
for discussions and insightful comments.

\end{document}